\documentclass[a4paper]{article}
\usepackage[pdftex]{graphicx}
\usepackage{INTERSPEECH2014,amssymb,amsmath,amsfonts,epsfig}
\usepackage{booktabs}
\usepackage{anyfontsize}
\usepackage{acronym}
\usepackage{perpage} 
\usepackage{url}
\newacro{AEC}{acoustic echo canceler}
\newacro{RAEC}{robust acoustic echo canceler}
\newacro{DTP}{double-talk probability}
\newacro{RES}{residual echo suppressor}
\newacro{NS}{noise suppressor}
\newacro{SER}{speech-to-echo ratio}
\newacro{SNR}{speech-to-noise ratio}
\newacro{STFT}{short time Fourier transform}
\newacro{NPE}{noise power estimator}
\newacro{RPE}{residual power estimator}
\newacro{GA}{genetic algorithm}
\newacro{ERN}{error recovery nonlinearity}
\newacro{MSE}{mean-squared error}
\newacro{MMSE}{minimum mean-squared error}
\newacro{SPP}{speech presence probability}
\newacro{LSA}{log-spectral amplitude}
\newacro{DD}{decision-directed}
\newacro{VAD}{voice activity detector}
\newacro{PAR}{phone accuracy rate}
\newacro{MOS}{mean opinion score}
\def\Vhrulefill{\leavevmode\leaders\hrule height 0.7ex depth
  \dimexpr0.4pt-0.7ex\hfill\kern0pt}

\ninept
\def\reg{{\rm\ooalign{\hfil
     \raise.07ex\hbox{\scriptsize R}\hfil\crcr\mathhexbox20D}}}

\title{Design and Optimization of a Speech Recognition Front-End for \\Distant-Talking Control of a Music Playback Device}

\makeatletter
\def\name#1{\gdef\@name{#1\\}}
\makeatother \name{{\em Ramin Pichevar, Jason Wung, Daniele Giacobello, and Joshua Atkins}}
\address{Beats Electronics, LLC, 1601 Cloverfield Blvd., Santa Monica, CA 90404, USA\\
{\small \tt ramin.pichevar@beatsbydre.com}}

\begin{document}
\maketitle
\begin{abstract}
This paper addresses the challenging scenario for the distant-talking control of a music playback device, a common portable speaker with four small loudspeakers in close proximity to one microphone. The user controls the device through voice, where the speech-to-music ratio can be as low as $-30$~dB during music playback. We propose a speech enhancement front-end that relies on known robust methods for echo cancellation, double-talk detection, and noise suppression, as well as a novel adaptive \emph{quasi}-binary mask that is well suited for speech recognition. The optimization of the system is then formulated as a large scale nonlinear programming problem where the recognition rate is maximized and the optimal values for the system parameters are found through a genetic algorithm. We validate our methodology by testing over the TIMIT database for different music playback levels and noise types. Finally, we show that the proposed front-end allows a natural interaction with the device for limited-vocabulary voice commands.
\end{abstract}
\let\thefootnote\relax\footnote{The authors thank Stephen Nimick for recording the voice commands used in the experimental evaluation.}

\section{Introduction}
The human interaction paradigm with music playback devices has seen a dramatic shift as devices get smaller and more portable. Well-established interaction media such as remote controls are no longer adequate. Automatic speech recognition (ASR) interfaces offer a natural solution to this problem, where these devices are typically used in hands-busy, mobility-required scenarios \cite{helander1997handbook}. Performing ASR on these small devices are highly challenging due to the music playback itself, the environmental noise, and the general environmental acoustics, e.g., reverberation \cite{deng2013speech}. In particular, due to the severe degradation of the input signal, the ASR performance drops significantly when the distance between the user and the microphone increases \cite{wolfel2009distant}. In the past decade, the literature on distant-talking speech interfaces provided several solutions to the problem, e.g., the DICIT project \cite{marquardt2009natural}. However, to the authors' knowledge, the available solutions rely heavily on large microphone arrays \cite{Seltzer2003}, which may be infeasible for handheld portable device.

In this work, we present a robust front-end speech enhancement and ASR solution for a single-microphone limited-vocabulary system during continuous monaural music playback. In contrast to previous studies, the microphone in our system is placed in close proximity to the loudspeakers, and the voice command still needs to be recognized at a very low \ac{SER} while the music is playing.

The front-end algorithm design effort can be divided in two parts. Firstly, we tailor known double-talk robust solutions for echo cancellation and speech enhancement to retrieve a clean estimate of the command \cite{Wada:2009gj,Wung:2011it,Wada:2012gt}. Secondly, we propose a novel noise reduction method, where we combine a traditional \ac{MMSE} speech enhancement approach \cite{Ephraim:1985bm} with an estimate of the ideal binary mask \cite{Hartmann2013}. The parameters of the algorithm are tuned for maximum recognition rate by casting the tuning problem as a nonlinear program, solved efficiently through a genetic algorithm (GA) \cite{goldberg1989genetic}. A similar approach was used in \cite{Giacobello2014,giacobello2013results} to maximize the objective perceptual quality of a speech enhancement system for full-duplex communication. The training and evaluation corpora are generated through a synthetic mixture of clean speech (from the TIMIT database \cite{garofolo1993timit}) and music, both convolved with separate impulse responses, and further mixed with a background noise to cover as many deployment scenarios as possible. The acoustic models of the ASR are trained by the front-end enhanced speech, an effective way to learn and exploit the typical distortions of the system itself \cite{huang2008effective}.



The paper is organized as follows. In Section~\ref{sec:algorithm}, we describe the speech enhancement algorithm and outline the parameters to be tuned. The tuning by nonlinear optimization of these parameters is presented in Section \ref{sec:tuning}. The experimental results in Section~\ref{sec:speechTIMIT} are divided in two parts. Firstly, we present the results of the training and evaluation of the front-end and acoustic models using the TIMIT database. Secondly, we change the language model and implement our ASR system for a limited vocabulary command recognizer in very adverse conditions. In Section \ref{sec:conclusions}, we conclude our work.

\section{Speech Enhancement System}
\label{sec:algorithm} 
\label{ssec:RAEC}
\begin{figure}[t]
  \centering
  \includegraphics[width= 1\columnwidth]{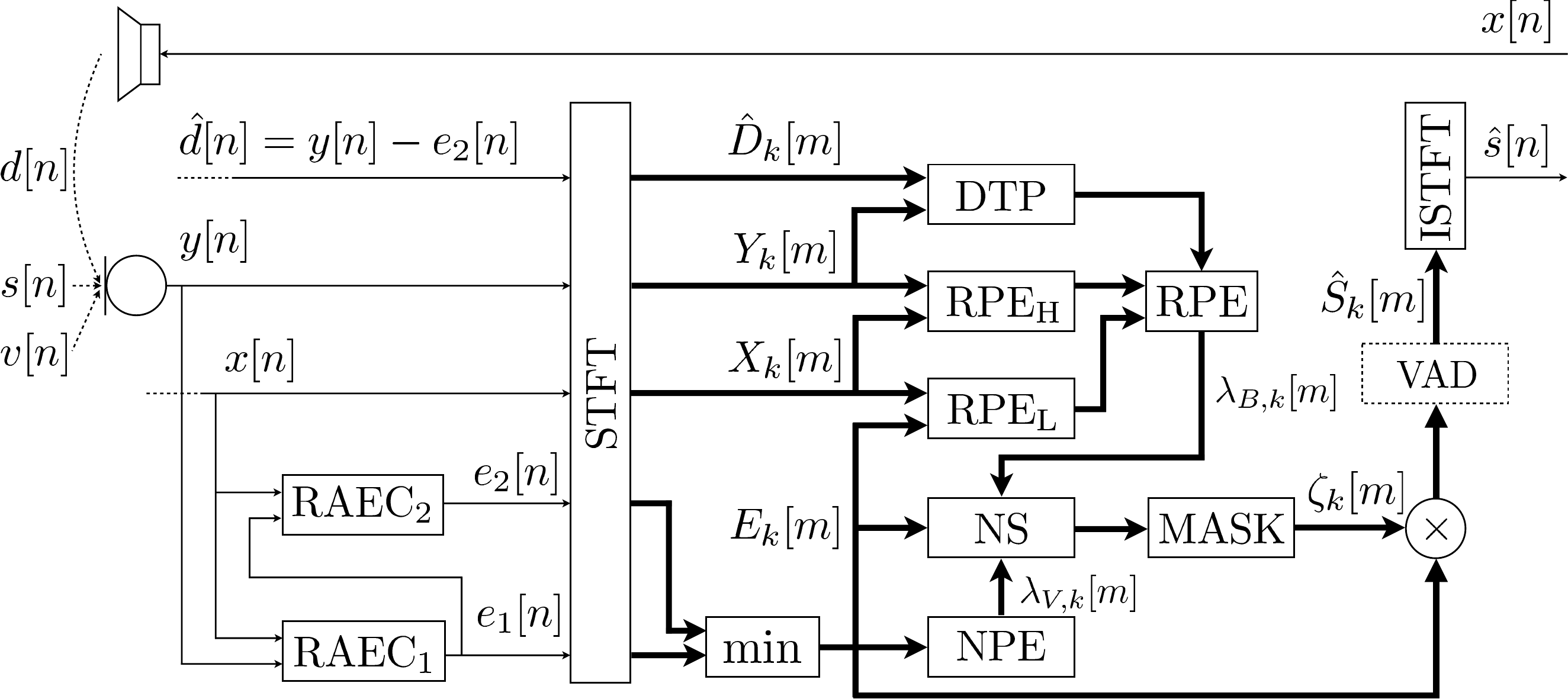}
  \caption{A block diagram of the speech enhancement system.}\vspace{-0.15in}
  \label{fig:wholeSystem}
\end{figure}
Let $y[n]$ be the near-end microphone signal, which consists of the near-end speech $s[n]$ and noise $v[n]$ mixed with the acoustic echo $d[n] = h[n] * x[n]$, where $h[n]$ is the impulse response of the system, $x[n]$ is the far-end reference signal, and $*$ is the convolution operator. The overall block diagram of the speech enhancement algorithm is shown in Figure~\ref{fig:wholeSystem}, which consists of two \acp{RAEC}, a \ac{DTP} estimator, two \acp{RPE}, a \ac{NPE}, a \ac{NS}, and a \ac{VAD}.

\subsection{Robust Acoustic Echo Canceler}

Since strong near-end interference may corrupt the error signal of the
\ac{AEC} and cause the adaptive filter to diverge, the \ac{RAEC}
system \cite{Wada:2009gj, Wada:2012gt} is used, where the error recovery nonlinearity
and robust adaptive step-size control allows for continuous tracking
of the echo path during double talk.
To reduce the delay of the frequency-domain adaptive filter
\cite{Shynk:1992bg}, the multi-delay adaptive filter structure
\cite{Soo:1990cx} is used.
A cascaded structure similar to the system approach of
\cite{Wung:2011it} is used:  the output of the first \ac{RAEC} is
fed to the input of the second \ac{RAEC}, which is different from the
original system approach in \cite{Wung:2011it} where the input to the
second \ac{RAEC} is still the microphone signal (a parallel structure
instead of the cascaded structure used in this work).

The tuning parameters for each of the \acp{RAEC} consist of the frame
size $N_\text{AEC}$, the number of partitioned blocks $M_\text{AEC}$,
the number of iterations $N_\text{iter}$,
the step-size $\mu_\text{AEC}$, the tuning parameter
$\gamma_\text{AEC}$ for the robust adaptive step-size, and the
smoothing factor $\alpha_\text{AEC}$ for the power spectral density estimation.

\subsection{Residual Echo Power Estimator}
\label{ssec:RES}

Since the \ac{AEC} cannot cancel all the echo signal due to modeling
mismatch, further enhancement from the \ac{RES} is required to
improve the voice quality. 
A coherence based method similar to \cite{Enzner:2002id,
  StefanGoetze:2005wf} is used for the \ac{RPE}, and a modified version of
the \ac{DTP} estimator similar to \cite{Tashev:2012et} is used for a
more accurate estimate of the residual echo power.
As shown in Figure~\ref{fig:wholeSystem}, the \ac{DTP} estimator differs from that in \cite{Tashev:2012et} since the coherence is calculated between the \ac{RAEC} estimated echo signal $\hat{d}$ and the microphone signal $y$ rather than between the loudspeaker signal $x$ and the microphone signal $y$. This is possible since the estimated echo signal $\hat{d}$ can be reliably obtained even during double talk due to the \textit{robust} echo path tracking performance of the \ac{RAEC}.

In this work, we propose to estimate the residual echo power by
utilizing the output of the double talk probability estimator.
Ideally, when the double-talk probability is high, the level of
residual echo power estimate should be low so as to not distort the
near-end speech when suppressing the residual echo.
On the other hand, when the double-talk probability is low, the level
of residual echo power estimate should be high to suppress as much residual echo
as possible.
The high level residual echo power $\lambda_{B_\text{H},k}$
 is estimated based on the coherence of the
microphone signal $Y_k$ and the reference signal $X_k$, while the low
level residual echo power $\lambda_{B_\text{L},k}$
 is estimated based on the coherence of the error
signal $E_k$ and the reference signal $X_k$.
Finally, the residual echo power $\lambda_{B,k}$ 
is estimated by utilizing the double-talk probability estimate $P_k^\text{DT}$
obtained from \ac{DTP} to combine
$\lambda_{B_\text{H},k}$ and $\lambda_{B_\text{L},k}$:
\begin{equation}
\label{eq:newRPE}
  \lambda_{B,k}[m] = (1-[m]P_k^\text{DT}[m])\lambda_{B_\text{H},k}[m] + P_k^\text{DT}[m]\lambda_{B_\text{L},k}[m],
\end{equation}
where $k$ is the frequency bin and $m$ time frame.

The tuning parameters for the \ac{DTP} consists of the transition
probabilities $a_{01}$, $a_{10}$, $b_{01}$, and $b_{10}$, the
 smoothing factors $\alpha_\text{DTP}$ and $\beta_\text{DTP}$, the frequency bin
 range $[k_\text{begin}, k_\text{end}]$, the frame duration
 $T_\text{DTP}$, and
the adaptation time constants $\tau$.
The tuning parameters for the \ac{RPE} consist of the numbers of
partitions $M_{\text{RPE}_\text{H}}$ and $M_{\text{RPE}_\text{L}}$ to
calculate the coherence and the smoothing factors
$\alpha_{\text{RPE}_\text{H}}$ and $\alpha_{\text{RPE}_\text{L}}$ for
the power spectral density estimation.

\subsection{Noise Suppressor}
\label{ssec:NS}


In this work, we combine \ac{RPE} and \ac{NPE} for residual echo
and noise suppression using a single noise suppressor, as shown in Figure~\ref{fig:wholeSystem}.
The low complexity \ac{MMSE} noise power estimator
\cite{Gerkmann:2012io} is used for the \ac{NPE}, and the Ephraim and
Malah \ac{LSA} estimator \cite{Ephraim:1985bm} is used for
the combined residual echo and noise suppression:
\begin{equation}
G_k^\text{LSA}[m] = \frac{\xi_k[m]}{1+\xi_k[m]} \mathrm{exp}\bigg( \frac{1}{2}
  \int_\frac{\xi_k[m] \gamma_k[m]}{1+\xi_k[m]}^{\infty} \frac{e^{-t}}{t}
  \,\mathrm{d}t\bigg).\label{eq:LSA}
  \end{equation}
 The estimation of the \textit{a priori} \ac{SNR} $\xi_k$ is done using the
\ac{DD} approach \cite{Ephraim:1984jc}:
\begin{align}\nonumber
\xi_k[m] &= \alpha_\text{DD} \frac{|\hat{S}_k[m-1]|^2}{\lambda_{V,k}[m] +\lambda_{B,k}[m]} \\ \nonumber
&\quad+(1-\alpha_\text{DD})\mathrm{max} \{\gamma_k[m] -1, 0\},
\end{align}
where
\begin{equation}
\gamma_k[m] = \lambda_{E,k}[m]/(\lambda_{V,k}[m] + \lambda_{B,k}[m])\nonumber
\end{equation}
and $\lambda_{E,k}$, $\lambda_{V,k}$, and $\lambda_{B,k}$ are the residual error signal power, the noise power, and residual echo power respectively. 

The tuning parameters of the \ac{NPE} consist of the fixed \textit{a priori} \ac{SNR} $\xi_{H_1}$, the threshold $P_\text{TH}$, and the
smoothing factors $\alpha_P$ and $\alpha_\text{NPE}$
The tuning parameters of the the \ac{NS}
consist of the smoothing factor for the \ac{SNR} estimator
$\alpha_\text{DD}$.
\subsection{Generation of Speech Enhancement Mask}

\begin{figure}[t]
\centering
\includegraphics[width=\columnwidth]{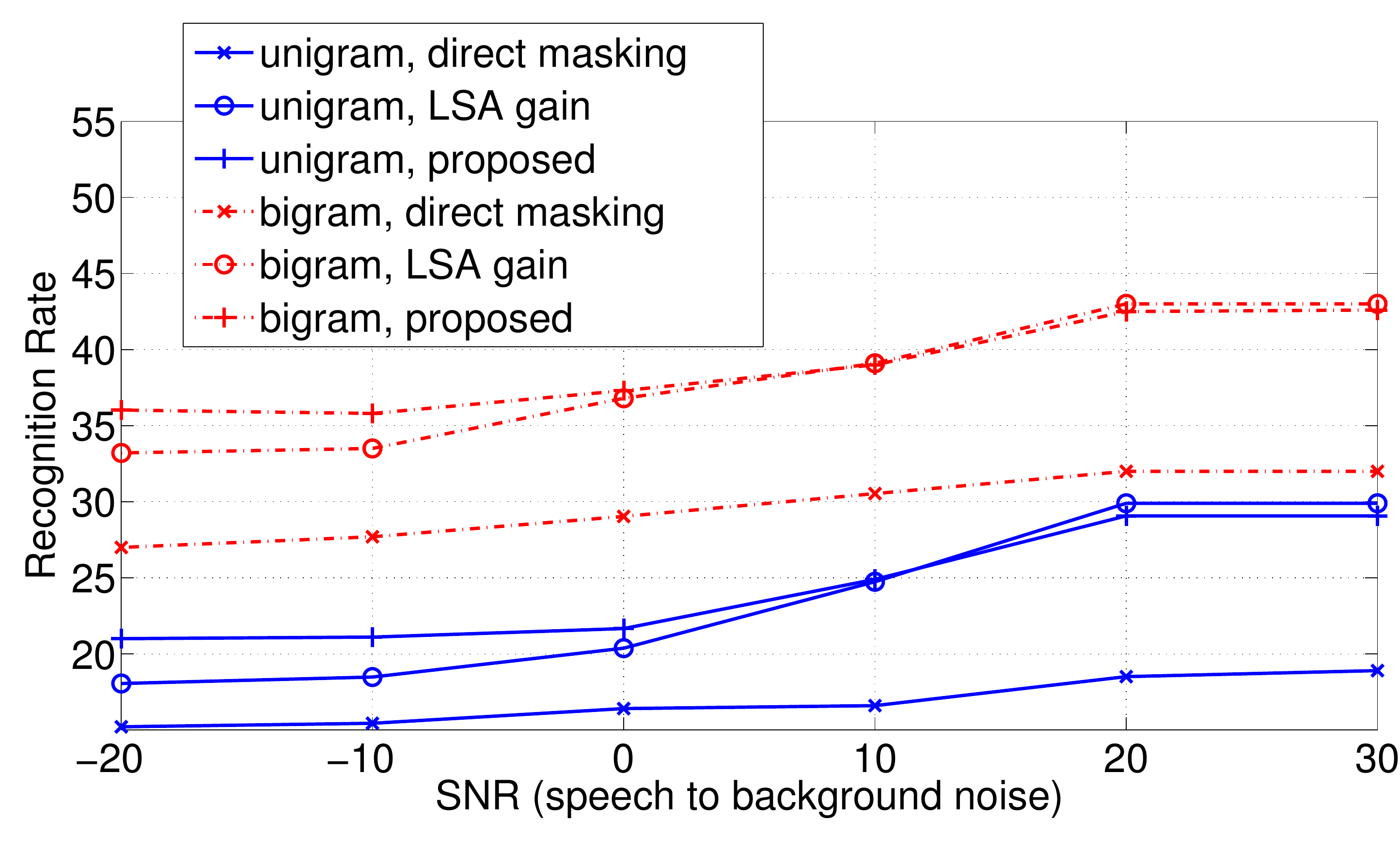}
\caption{Comparison of recognition rates on the noisy TIMIT database for our proposed  direct masking with direct binary masking and MMSE-based LSA gain for different speech to background SNRs and a constant SER of $-20$~dB. Recognition rate for the bigram model with an ideal binary mask is around 65\% throughout the SNR range.}
\vspace{-0.15in}
\label{fig_direct}
\end{figure}

It has been recently shown that the speech recognition accuracy in noisy condition can be greatly improved by direct binary masking \cite{Hartmann2013} when compared to marginalization \cite{Cooke2001} or spectral reconstruction \cite{Raj2004}. Given our application scenario, we propose to combine the direct masking approach, particularly effective at low overall \acp{SNR}, with the \ac{NS} output mask $G_k^\text{LSA}$, as shown in Figure~\ref{fig:wholeSystem}. In particular, we exploit the estimated bin-based \emph{a priori} \ac{SNR} $\xi_k$ to determine the type of masking to be applied to the spectrum. However, given than an accurate estimation of the binary mask is very difficult for very low \acp{SNR}, we elect to use the \ac{LSA} estimated gain for those cases.
Our masking then becomes:
\begin{align}\nonumber
\zeta_k[m] &= 
\begin{cases}
[(1 - G_\text{min})G_k^\text{LSA}[m] + G_\text{min}], &\xi_k[m]\le\theta_1, \\
 \frac{\alpha }{2},                              & \theta_1<\xi_k[m]<\theta_2,\\
 \frac{2+\alpha}{2},                                &\xi_k[m]\ge\theta_2,  
\end{cases}
\end{align}
where $G_\text{{min}}$ is the minimum suppression gain \cite{giacobello2013results}, and the output is then:
\begin{equation}\label{eq:mask} 
  \hat{S}_k[m]  = \zeta_k[m]   E_k[m].
\end{equation}
In Figure~\ref{fig_direct}, we provide some data to justify our particular choice of masking. We compare  three different speech enhancement methods presented in this section for unigram and bigram language models \cite{Young2000}. In the direct masking, $\xi_k[m]$ is mapped directly to a constant threshold to generate the binary decision. It can be seen that our proposed method outperforms conventional methods at lower SNRs.

The tuning parameters for the direct masking consist of
the minimum gain $G_\text{min}$, the thresholds $\theta_1$ and $\theta_2$, and a tuning parameter $\alpha$.


\section{The Tuning Problem}\label{sec:tuning}
The tuning problem can be formalized as an optimization problem. In our case, the objective function to maximize is the ASR recognition rate $ {\bf R} \left(\hat{s}[n]\right)$, where $\hat{s}[n]$ is the processed speech, i.e., the output of the speech enhancement system. To restrict the search region, we can impose inequality constraints on the variables that simply determine lower and upper bounds limit for the components of the solution vector. Our optimization problem then becomes:
\begin{equation}
\begin{aligned}
& \underset{}{\text{maximize}}
& &  {\bf R} \left(\hat{s}[n, {\bf p}]\right) \\
& \text{subject to}
& & {\bf U} \leq {\bf p} \leq {\bf L},
\label{eq:maxmos}
\end{aligned}
\end{equation}
where ${\bf p}$ is now the vector of the parameters that need tuning, $\hat{s}[n, {\bf p}]$ is the speech enhancement system output obtained with these parameters, and ${\bf L}$ and ${\bf U}$ represent, respectively, lower and upper bounds for the values each variable. The basic concept of a GA is to apply genetic operators, such as \emph{mutation} and \emph{crossover}, to evolve a set of $M$ solutions, or \emph{population}, ${\mathbf \Pi}^{(k)} = \{ {\bf p}_{m}^{(k)}, m = 1, \ldots, M \} $ in order to find the solution that maximizes the cost function \cite{goldberg1989genetic,duda2012pattern}. This procedure begins with a randomly chosen population ${\mathbf \Pi}^{(0)} $ in the space of the feasible values $\left[{\bf L} ,{\bf U}\right]$, and it is repeated until a halting criterion is reached after $K$ iterations.
The set of parameters  ${\bf p}_{m}^{(K)}  \in {\mathbf \Pi}^{(K)}$ that maximizes the cost function will be our estimate:
\begin{equation}
 \label{eq:final}
 \hat{{\bf p}} = \underset{{{\bf p}_{m}^{(K)}  \in {\mathbf \Pi}^{(K)}}}{\mathrm{arg\,max}} {\bf R} \left(\hat{s}[n,{\bf p}_{m}^{(K)}]\right).
\end{equation}

\section{Experimental Results}\label{sec:speechTIMIT}
In this section, we present the results from our designed speech enhancement front-end with the tuned parameters using the optimization method presented in Section \ref{sec:tuning}. In order to obtain the set of parameters that maximize the recognition rate, we optimized and tuned the system on a noisy TIMIT database. The set of tuned parameters will then be used in the ASR front-end for the distant-talking limited-vocabulary control of our music playback device as shown in Figure~\ref{fig_setup}.

\subsection{Speech Recognition on TIMIT}

\subsubsection{Noisy Database}
The database was generated by simulating the interaction between the user and the playback device. In this scenario, music is played from a four-loudspeaker portable device with an embedded microphone, placed roughly one centimeter away from the closest loudspeaker, and the user is uttering speech in a reverberant environment during continuous music playback. The microphone signal $y[n]$ was then generated according to:
\begin{equation}
y[n]=s[n]+\sigma_1 d[n]+\sigma_2 v_2[n]+\sigma_3 v_3[n],\nonumber
\end{equation}
which consisted of the speech $s[n]$, the acoustic echo from the music $d[n]$, the background noise $v_2[n]$ (babble, factory, and music), and a pink noise introduced to simulate a mild broadband constant electrical noise and electromagnetic radiations $v_3[n]$. For each file in the TIMIT database, the SER and SNR were chosen from uniform distributions ranging from $-15$~dB to $-10$~dB and from $-10$~dB to $10$~dB, respectively. We used $12$ impulse responses in the simulation, randomly picked and normalized to unitary energy. The values of $\sigma_1$ and $\sigma_2$ were calculated based on SER and SNR, and we set $\sigma_3=0.1$. The music sound, $d[n]$, was randomly selected from five different music tracks of different genres with random starting points.

\subsubsection{Training of the Speech Recognizer}

We used the HTK toolkit \cite{Young2000} to train an acoustic model on the noisy TIMIT database composed of 61 phones \cite{lopes2011}. A set of 13 Mel-frequency cepstral coefficients (MFCCs) with their first and second derivatives, for a total of 39 coefficients, are generated and used as features for our experimental analysis.  We normalized the variance and mean of the MFCCs, as suggested in \cite{Hartmann2013} for properly applying the direct masking. We used 5-state HMMs with a 8-mixture GMM for each phone. We trained our HMMs with the noisy speech processed by our front-end. 
\begin{figure}[t]
\centering
\includegraphics[width=\columnwidth]{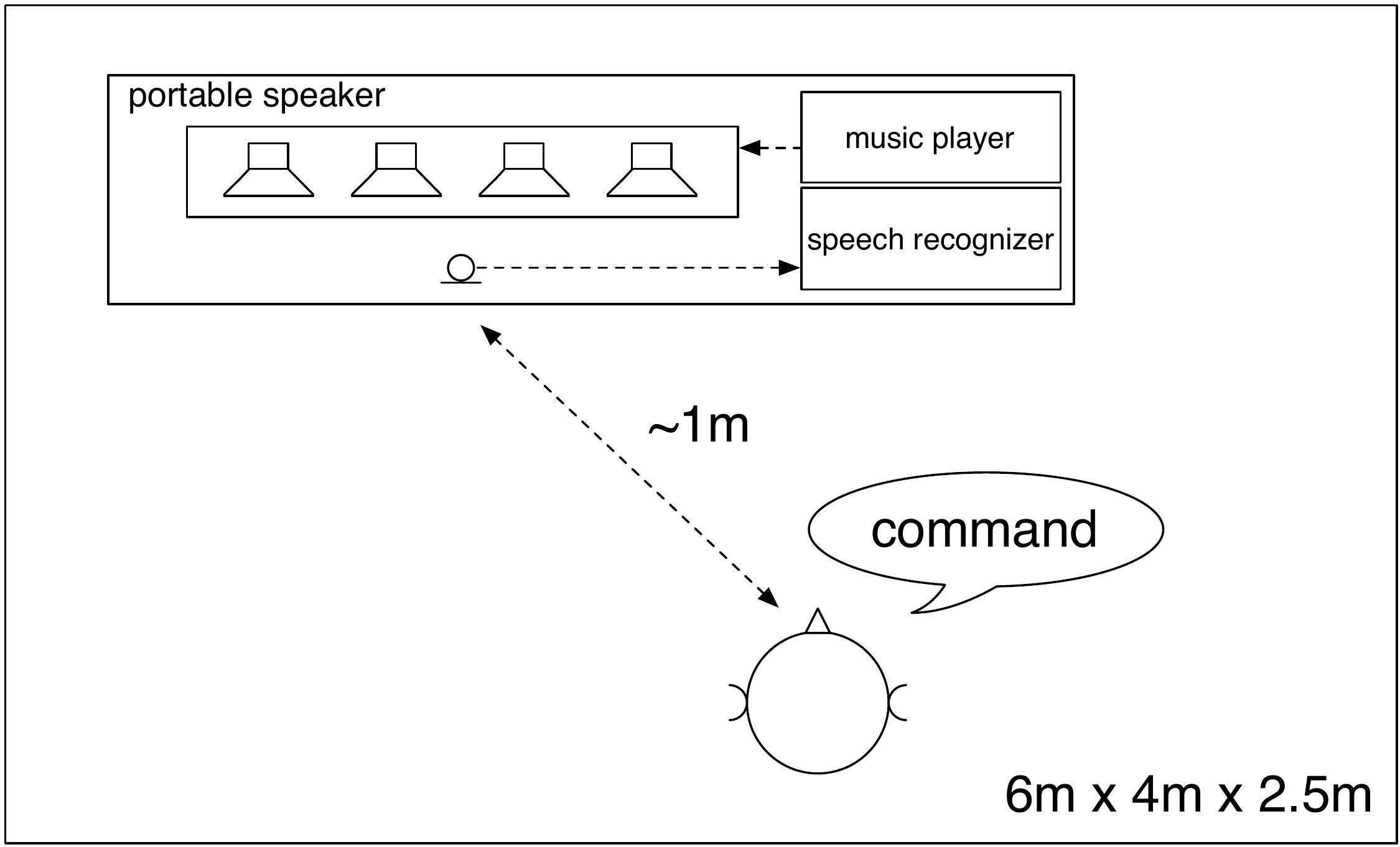}
\caption{Experimental setup for voice recording.}
\label{fig_setup}
\vspace{-0.15in}
\end{figure}
\setcounter{footnote}{0}
\subsubsection{Recognition of the noisy TIMIT database}
Once we obtained the HMMs in the acoustic model, we optimized the parameters of our front-end. We casted the problem as discussed in Section \ref{sec:tuning}. For initial population, we chose a set of fairly well manually optimized parameters and reasonable bounds that allows us to use only three generations to reach convergence. The genetic algorithm had a population of $M=40$ possible candidates, and the best $N=10$ were migrated to the next generation. These values were chosen empirically by balancing the complexity and the accuracy of the results. The \ac{PAR} using a bigram model increased from 35\% to 40\% after our optimization on the training data, proving the validity of our procedure.

In order to provide a fair comparison, we also tuned the parameters to maximize the \ac{MOS} using the Perceptual Objective Listening Quality Assessment (POLQA) \cite{polqa}, as done in \cite{Giacobello2014}, through the same GA setup and the same noisy TIMIT database. To assess the performance of our tuning method, we tested on data not used in the training by creating a second simulated noisy TIMIT database with different conditions. Results are shown in Table~\ref{tbl:TIMIT} for different types of noise. The SER and SNR were again chosen from uniform distributions ranging from $-15$~dB to $-10$~dB and from $-10$~dB to $10$~dB, respectively. The ``mix'' noise was picked randomly from the babble, music, or factory noise. In the case of music, noisy files were generated from a set of tracks from different genres at different start points. When the front-end speech enhancer was not used, the \ac{PAR} dropped to 10.1\% (unigram) and 15.7\% (bigram) for the noisy signal. Although used in a different setup, the results obtained with the proposed method compare favorably to some prior results \cite{Herbordt20051,Reuven2007}, where authors investigated joint echo cancellation and speech enhancement at higher SERs and SNRs.
\begin{table}[t]
\begin{center}
\caption{Phone Accuracy (\%) for the noisy TIMIT database .}
\label{tbl:TIMIT}
\scalebox{0.85}{
\begin{tabular}{l|cc|cc|cc|cc}
\hline
noise &\multicolumn{2}{c|}{mix} &\multicolumn{2}{c|}{babble} &\multicolumn{2}{c|}{music} &\multicolumn{2}{c}{factory} \\
\hline
   model & uni.    & bi. &uni.   &bi.  & uni.   &bi. & uni.   &bi.\\
\hline
   ASR & \textbf{22.7} & \textbf{37.4} &\textbf{22.4} &\textbf{37.0} &\textbf{22.2} &\textbf{36.3} &\textbf{21.6} &\textbf{36.5}\\
   POLQA &21.7 &35.7 &21.6 &35.6 &21.1 &35.3 &21.4 &35.5\\
\hline
    \end{tabular}
}
\end{center}
\vspace{-0.15in}
\end{table}

\subsection{Limited Vocabulary Speech Recognition}\label{sec:speechLimited}
We used the set of tuned parameters and the HMMs obtained from our analysis on the TIMIT database to study the feasibility of speech recognition on limited vocabulary in extremely challenging conditions.
\subsubsection{Recognition of limited-size Vocabulary Speech}
\begin{figure}[t]
\begin{center}
\includegraphics[width=\columnwidth]{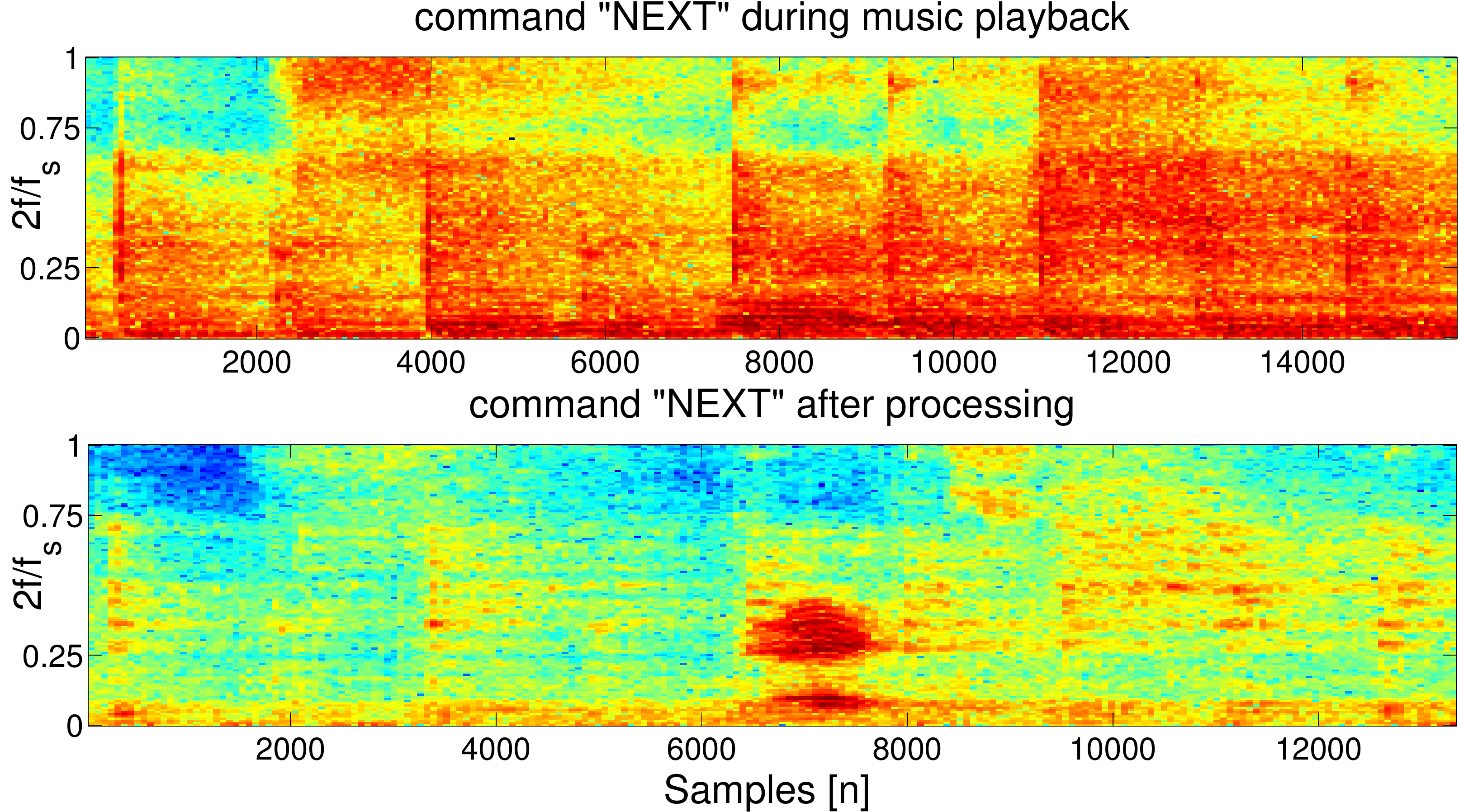}
\caption{Spectrogram of the original mixture and cleaned speech for an instance of  the ``NEXT" command.} \label{fig:next}
\vspace{-0.15in}
\end{center}
\end{figure}

We used the system to recognize four commands: ``PLAY'', ``NEXT'', ``BACK'', and ``PAUSE''. The commands were generated by changing the TIMIT language model accordingly. As shown in Figure \ref{fig:wholeSystem}, we used a standard \ac{VAD}, applied on a frame-by-frame basis, after the direct masking to isolate the commands \cite{Ramirez2007,sohn1999}:
\begin{equation} \label{eq:VAD}
\sum_k\left[\frac{\gamma_k \xi_k}{1+\xi_k}-\text{log}(1+\xi_k)\right]>\eta,
\end{equation}
where $\xi_k$ and $\gamma_k$ are the \textit{a priori} and \textit{a posteriori} SNRs and $\eta$ is a fixed threshold. Figure~\ref{fig:next} shows an example of a noisy command before and after processing. The command is not audible to human listeners before processing, while the speech structure is well preserved after processing.

\subsubsection{Recording of Real-World Commands}
We used eight subjects (male/female, native/non-native) who uttered the command list at a distance of around 1m from the microphone of the \emph{Beats Pill}\texttrademark\ portable speaker while music was playing. We used four different music tracks in the echo path, where the starting point of the track was chosen randomly. Subjects uttered the following commands towards the speakers: ``PLAY'', ``NEXT'', ``BACK'', ``PAUSE'' (as shown in Figure~\ref{fig_setup}). The playback level for the experiments was set to three different levels of $95$~dB SPL, $90$~dB SPL, and $85$~dB SPL. We estimated the range of SER for the different setups to be approximately equal to $-35$ to $-30$~dB, $-30$ to $-25$~dB, and $-25$ to $-20$~dB for the three levels, respectively. The estimation of the SERs were made possible thanks to a \emph{lavalier} microphone that recorded the near-end speech. Note that the SERs in the experiments are lower than the SERs used in the simulation, which validates the generalization of the tuning methodology. Recognition rates are given in Table~\ref{tbl:limited} at different SER levels. Also in this case, we compared with the set of parameters obtained by optimization through POLQA \cite{Giacobello2014}. The results clearly show that our proposed tuning based on ASR maximization outperforms the POLQA-based tuning. The difference in performance seems to derive from the POLQA optimization being less aggressive on noise in order to preserve speech quality. More noise in the processed files translates into worse performance of the speech recognizer and the VAD. As a reference, when our speech enhancement front end was not used, the average recognition rate was 25\% over all commands (coin toss) in the lowest SER setup.
\begin{table}[t]
\caption{Command Accuracy (\%) for different commands at different SERs.} \label{tbl:limited}
\begin{center}
\scalebox{0.85}{
\begin{tabular}{l |c c |c c |c c }
\hline
 SER (dB) &\multicolumn{2}{c|}{$-35 \sim -30$}   &\multicolumn{2}{c|}{$-30 \sim -25$}  &\multicolumn{2}{c}{$-25 \sim -20$}  \\
\hline
   params.                   & ASR    & POLQA           &ASR   &POLQA            & ASR   &POLQA\\
\hline
   BACK                             & 73       &47                   &83      &50                     &90       &53 \\
   NEXT                             & 70       &50                   &90       &57                   &90       &63\\
   PLAY                              & 80      &67                    &94      &80                  &96        &83 \\
   PAUSE                           & 76      &50                    &87       &57                 &87        &60\\
\hline
    \end{tabular}
}
\end{center}
\vspace{-0.15in}
\end{table}



\section{Conclusion}\label{sec:conclusions}
We proposed a robust ASR front-end and a related tuning methodology. The proposed speech enhancement front-end consists of a cascaded robust \ac{AEC}, a residual echo power estimator based on a double-talk probability estimator, and a novel \textit{quasi}-binary masking that utilizes the classical MMSE-based method at very low SNRs. The tuning improves the speech recognition rate substantially on the TIMIT database. The optimized front-end is then tested in realistic environments for the remote control of a music playback device with a limited-sized command dictionary. The result shows a fairly high recognition rate for voice commands at a speech-to-music ratio as low as $-35$~dB, scenarios hardly seen through the literature.






%


\end{document}